\documentclass[aps,prl,twocolumn,groupedaddress,showpacs]{revtex4}

\usepackage{amsmath}
\usepackage{graphicx}
\usepackage{hyperref}
\newcommand{\Matrr}[2]
{\left[ \begin{array}{c} #1 \\ #2 \end{array} \right]}

\begin{document}

\title{Persistent currents through a quantum impurity: Protection 
through integrability}

\author{Johan Nilsson,$^1$ Hans-Peter Eckle$^2$ and Henrik 
Johannesson$^3$}
\affiliation{$\mbox{}^1$Department of Physics, Boston University, 
Boston, MA 02215}
\affiliation{$\mbox{}^2$Advanced Materials Science, University of Ulm, 
D 89069 Ulm, Germany}
\affiliation{$\mbox{}^3$Department of Physics,
              G\"oteborg University,
              SE-412 96 G\"oteborg, Sweden}

\date{\today}

\begin{abstract}

We consider an integrable model of a one-dimensional mesoscopic ring 
with the conduction electrons coupled
by a spin exchange to a magnetic impurity. A 
symmetry analysis based on a {\em Bethe Ansatz}
solution of the model reveals that the current is insensitive to the 
presence of the impurity.
We argue that this is true for {\em any} integrable impurity-electron 
interaction, independent of choice of
physical parameters or couplings. We propose a simple physical
picture of how the persistent current gets protected by integrability.

\end{abstract}

\maketitle

The physics of quantum impurities has become an important chapter in 
the evolving theory of strongly correlated matter.
The reasons are several: First, quantum impurity problems arguably 
represent the simplest settings in which to
analyze various aspects of electron correlations. A case in point is 
the Kondo effect where
a magnetic impurity induces an effective electron-electron interaction 
that increases as the energy scale is
lowered. Various extensions of the original problem have opened up 
entire new fields of investigations, from the
modeling of correlated transport in DNA molecules, to novel scenarios 
for non-Fermi liquid behaviors
\cite{40years}.
Secondly, quantum impurity problems are often tractable by exact 
analytical methods, most prominently the
{\em Bethe Ansatz} technique which exploits the integrability of the 
paradigmatic Kondo- and Anderson models
\cite{baKondo}.
This has proven immensely
useful, with the exact results serving as ''benchmarks'' for more 
conventional, numerical or perturbative methods.

Most importantly, progress in experiments on mesoscopic and 
nanoscale systems now enables controlled studies of a
single quantum impurity interacting with conduction electrons. In 
groundbreaking experiments in the late 90s
\cite{Goldhaber},
semiconductor quantum dots, connected capacitively to a gate and via 
tunnel junctions to electrodes, were
shown to exhibit a tunable Kondo effect: Below a characteristic 
temperature $T_K$, a single electron occupying the
highest spin-degenerate level of the dot forms a spin-singlet with 
electrons in the leads, producing a Kondo resonance
at the Fermi level. This and subsequent developments \cite{MesoReview} 
have turned quantum impurity physics into an
essential piece also of modern nanoscience.

An interesting problem in this context is how the charge persistent current 
(PC) in a mesoscopic ring coupled to a
quantum dot is affected by a Kondo resonance. A PC is the equilibrium 
response to a magnetic Aharonov-Bohm
flux piercing the ring \cite{Viefers}. It requires for its existence 
that an electron maintains its phase coherence
while encircling the ring, and
is thus expected to be sensitive to scattering off the quantum dot. In 
a previous study \cite{EJS}, it was found that
the PC is amenable to a {\em Bethe Ansatz} analysis when the quantum 
dot is side-coupled to the ring.
For certain privileged
values of the flux, the problem was mapped onto the integrable 
one-dimensional (1D) Kondo model with a linearized dispersion.
Contrary to expectation it was found that the Kondo impurity that 
represents the dot has no effect on the persistent
current. While this result conforms with those of some other authors 
\cite{Zvyagin,Cho}, a well-controlled RG analysis \cite{AffleckSimon}
together with large-scale numerics \cite{Sorensen} 
strongly suggest that the PC in fact vanishes when the
ring is larger than the Kondo screening cloud 
(other related work includes Refs. \cite{AS2006,Aligia02}).
This raised doubts about the applicability of a {\em Bethe
Ansatz} approach \cite{AffleckSimonComment}. Having served for many 
years as a work horse in the study
of bulk quantum impurity physics, the 1D integrable Kondo model was 
now perceived to suffer from a difficulty when applied to
this particular problem: Its linear dispersion relation,
in addition to decoupling spin and charge degrees of freedom,
enforces
a non-standard procedure for extracting the PC from the finite-size 
spectrum. 
It was suggested that these features likely explain the failure to 
obtain an
effect from the impurity on the PC \cite{AffleckSimonComment}.

In an attempt to shed light on this
intriguing issue, we investigate, in this Letter, the influence of a 
local magnetic moment on the PC in a mesoscopic ring, using an 
integrable model with a a {\em non--linear} dispersion relation for the 
electrons. As in Ref. \cite{EJS}, the impurity is coupled to the ring 
in such a way that the ring is unaffected when the coupling to the 
impurity is switched off. 
Unlike the analysis in Ref. \cite{EJS}, however, we do {\em not} 
linearize the electronic spectrum, but keep the
parabolic dispersion of non--relativistic electrons since we want to 
expressly study its possible effect on the PC. Apart from the 
results implied by Refs. 
\cite{AffleckSimon,Sorensen}, the electronic band curvature is
not known to play any significant role in the physics of the Kondo 
effect. Nevertheless, a small number of studies
have addressed the question of nonlinearities in the electronic 
spectrum in the Kondo problem, with a particular
eye on how to preserve integrability of the model 
\cite{Schulz,
WangVoit,LiBares}. In what follows we shall draw
on some of the insights gained from these studies.

The basic building blocks that go into the construction of an 
integrable model are the two-particle scattering matrices $S_{ij}$
\cite{IntegrableReview}.
These have to satisfy the Yang-Baxter equation
\begin{equation} \label{YB}
S_{ij}S_{ik}S_{jk} = S_{jk}S_{ik}S_{ij},
\end{equation}
the hallmark of integrability \cite{IntegrableReview}.
Constructing the electron--impurity scattering matrix by the same 
procedure as for the ordinary Kondo model \cite{baKondo} but now with a 
quadratic spectrum necessitates for consistency the introduction of a 
local potential term in the 
Hamiltonian:
\begin{equation}
V_{c}(x)\propto
\bigl[\delta^\prime(x^{+})+\delta^\prime(x^{-}) \bigr] x/|x| ,
\label{eq:counter}
\end{equation}
with $x=0$ the location of the impurity.
Moreover, to satisfy the Yang--Baxter equations (\ref{YB}) for 
electron--impurity and electron--electron scattering, the electrons 
must
interact via a local interaction whose strength
is adapted to the Kondo coupling of the magnetic moment.
The inclusion of interacting electrons implies a dichotomy: attractive 
electron--electron interaction necessitates an antiferromagnetic Kondo 
coupling, while repulsive electron--electron interaction implies  
a ferromagnetic Kondo coupling.
We shall concentrate here on the latter case.
Since our interest is to study the consequences of a non--linear band 
structure in the framework of an integrable model, both, the
auxiliary potential in eq.\ (\ref{eq:counter}) and the dichotomy 
between electron--electron interaction and Kondo coupling, can be 
easily tolerated.
We note in passing that a mechanism leading to a ferromagnetic Kondo 
coupling in quantum dots has recently been suggested by Silvestrov and 
Imry \cite{si}.

The first-quantized Hamiltonian on a ring of circumference $L$ 
consistent with integrability as outlined above is given by \cite{LiBares}:
\begin{eqnarray}
H & = & \sum_{i} \bigl[ -\partial^2_{x_i}
+ (J\vec{\sigma}_{i} \cdot \vec{\sigma}_{0} + J')\delta(x_i) \bigr]
\nonumber \\
& &+ \sum_{i}V_{c}(x_{i})
+ \sum_{i<j} 2c \: \delta(x_j-x_i),
\label{eq:schrodinger_spec}
\end{eqnarray}
where $2J=-c<0$ and $J'=-J$ are required by integrability.
The integrability of the model allows for an exact solution, encoded by 
the
{\em Bethe Ansatz} equations (BAE)
\begin{subequations}
\label{eq:zdef}
\begin{multline}
\label{eq:zdef_a}
I_{j}/L = z_{c}(k_j) =
\frac{k_j}{2\pi} - \frac{1}{2\pi L}\sum_{\gamma = 1}^{N_s} 
\theta_{1/2}(k_j-\lambda_{\gamma})
\end{multline}
\begin{multline}
\label{eq:zdef_b}
J_{\gamma}/L = z_{s}(\lambda_{\gamma}) =
- \frac{1}{2\pi L} \biggl[ \, \sum_{j = 1}^{N_c}
\theta_{1/2}(\lambda_{\gamma}-k_j)  \\ 
- \sum_{\delta = 1}^{N_s} 
\theta_{1}(\lambda_{\gamma}-\lambda_{\delta})
+ \theta_{1/2}(\lambda_{\gamma}) \biggr].
\end{multline}
\end{subequations}
Here $k_j\, (\lambda_{\gamma})$ are rapidities of the charge (spin) 
degrees of freedom,
and
$\theta_{n}(x)\equiv -2\tan^{-1}(x/nc)$.
Note that, except for the last term on the right--hand side of 
eq.\ (\ref{eq:zdef_b}) 
these are the BAE for the
repulsive $\delta$-function Fermi gas \cite{Yang}.
This last term, however, encapsulates the contribution of the
localized
magnetic moment. As is well known \cite{Viefers}, an Aharonov-Bohm flux 
threading the ring is equivalent to imposing twisted boundary 
conditions and in our case adds a term proportional to the flux to
eq.~(\ref{eq:zdef_a}).
We shall include such a flux term into our analysis conveniently at a 
later stage.
The quantum numbers $I_j$ and $J_\gamma$ are integers or half--integers 
depending on the number of electrons,
$N_c = N_{\uparrow} + N_{\downarrow}$,
and the number of down--spin electrons, $N_s = N_{\downarrow}$.
Their maximal and minimal values
are $I^\pm$ and $J^\pm$, respectively. Thus
\begin{subequations}
\label{eq:Ndef}
\begin{align}
&N_{c} = I^{+}-I^{-}+1, \qquad &N_{s} = J^{+}-J^{-}+1, \\
&D_{c} = ( I^{+}+I^{-} )/2 , \quad &D_{s} = ( J^{+}+J^{-} ) / 2.
\end{align}
\end{subequations}
$N_c$ and $N_s$
constitute the numbers of particles in the charge and spin Fermi seas.
$D_c$ and $D_s$ are the numbers of electrons and down--spin electrons 
moved from the left to the right Fermi points of their
respective Fermi seas.

It is important to note that in contrast to the standard 1D Kondo model 
with a linearized spectrum, charge- and spin degrees of freedom
are coupled through the BAE in (\ref{eq:zdef}). This suggests that the 
presence of the magnetic impurity may now feed back on the charge 
sector and affect 
the PC. To find out whether this
happens requires a careful analysis, to be expounded in what follows.

Introducing the root density functions 
$\vec{\rho}=(\rho_c, \rho_s)^T$,
\begin{equation}
\rho_{c}(k) = \frac{\partial z_{c}(k)}{\partial k}, \qquad 
\rho_{s}(\lambda) = \frac{\partial z_{s}(\lambda)}{\partial \lambda},
\label{eq:rhodef}
\end{equation}
in the charge $(c)$ and spin $(s)$ sectors, we can employ
the well--known framework \cite{ferenc} for extracting the lowest order 
finite--size corrections to the ground--state energy in the 
thermodynamic limit.
These are the finite--size corrections which determine the PC.
Using the Euler--Maclaurin formula for converting sums into integrals, 
one can retain finite--size corrections (in principle to arbitrary 
order) when converting the BAE (\ref{eq:zdef}) into a set of 
inhomogeneous coupled linear integral equations for the root densities 
\cite{ferenc}.
The phase shifts of eqs.\ (\ref{eq:zdef}) translate, via 
\ (\ref{eq:rhodef}) 
and the Euler--Maclaurin formula, into the integral 
kernel and the inhomogeneity of
these integral equations, respectively.
Given that the phase shifts are odd functions of their respective 
arguments, the inhomogeneities and therefore the solutions of the 
integral equations 
also attain a certain definite symmetry.
Our analysis of the PC will eventually rely exclusively on exploiting 
this symmetry.
To achieve a finite--size energy expression correct to order $1/L$, we
introduce the integration limits ($k^{\pm}$,$\lambda^{\pm}$) 
by
\begin{equation}
L \, z_c(k^{\pm})       = I^{\pm} \pm 1/2 , \qquad
L \, z_s(\lambda^{\pm}) = J^{\pm} \pm 1/2 ,
\label{eq:kdef3}
\end{equation}
such that equations (\ref{eq:Ndef}) become
\begin{subequations}
\label{eq:functionsdef}
\begin{align}
&\frac{N_{c}}{L} =
\int_{k^{-}}^{k^{+}} d k \rho_{c}(k),  \qquad
\frac{N_{s}}{L} = \int_{\lambda^{-}}^{\lambda^{+}} d \lambda
\rho_{s}(\lambda),  \\
&\frac{D_{c}}{L} =
z_{c}(0) + \frac{1}{2} \biggl[
\int_{0}^{k^{-}} dk  +
\int_{0}^{k^{+}} dk \biggr] \rho_{c}(k) \\
&\frac{D_{s}}{L} =
z_{s}(0) + \frac{1}{2} \biggl[
\int_{0}^{\lambda^{-}} d \lambda
  +
\int_{0}^{\lambda^{+}} d \lambda
\biggr] \rho_{s}(\lambda),
\end{align}
\end{subequations}
where, from (\ref{eq:zdef})
\begin{subequations}
\label{eq:z0}
\begin{align}
&z_{c}(0) =
\frac{1}{2\pi}\int_{\lambda^-}^{\lambda^+}
\rho_{s}(\lambda)\theta_{1/2}(\lambda) d \lambda
\\
&z_{s}(0) = \frac{1}{2\pi} \biggl[ \int_{k^-}^{k^+} d k
\rho_{c}(k)\theta_{1/2}(k)
-\int_{\lambda^-}^{\lambda^+} d \lambda
\rho_{s}(\lambda)\theta_{1}(\lambda)  \biggr].
\end{align}
\end{subequations}
The formal solution of the integral equations can then be decomposed as
\begin{multline}
\vec{\rho}(k,\lambda) = \vec{\rho}_{\infty}(k,\lambda) +
\frac{1}{L}\vec{\rho}_d(k,\lambda) \, + \\
\frac{1}{24L^2} \biggl[
\frac{\vec{\rho}_{1}(k,\lambda | k^{\pm},\lambda^{\pm})}{\rho_c(k^+)} +
\frac{\vec{\rho}_{1}(-k,-\lambda | 
-k^{\mp},-\lambda^{\mp})}{\rho_c(k^-)}  \\
+ \frac{\vec{\rho}_{2}(k,\lambda | 
k^{\pm},\lambda^{\pm})}{\rho_s(\lambda^{+})} \: +
\frac{\vec{\rho}_{2}(-k,-\lambda | 
-k^{\mp},-\lambda^{\mp})}{\rho_s(\lambda^{-})} \biggr].
\label{eq:rhofinal}
\end{multline}
Here $k^\pm$ and $\lambda^\pm$ play the role of Fermi points of the 
charge and spin excitations.
The densities in (\ref{eq:rhofinal}) therefore depend on the numbers 
$I^\pm$ and $J^\pm$, or, via (\ref{eq:Ndef}), on the parameters $N_r$ 
and $D_r$ ($r=c,s$).
The form of (\ref{eq:rhofinal}) depends crucially on the fact that the 
charge rapidities $k_j$ enter with odd symmetry into the BAE 
(\ref{eq:zdef}).
The term $\vec{\rho}_{d}/L$ describes the finite-size contribution of 
the magnetic moment, with
$\vec{\rho}_{\infty}$, $\vec{\rho}_{d}$, $\vec{\rho}_{1}$ and 
$\vec{\rho}_{2}$
solving the integral equations with appropriate inhomogeneous parts.
In particular,
\begin{equation}
\begin{split}
\vec{\rho}_{\infty0} = \Matrr{1/2\pi}{0}
\qquad
\vec{\rho}_{d0} = \Matrr{0}{K_{1/2}(\lambda)} \qquad
\label{eq:rho0def}
\end{split}
\end{equation}
with the kernel
$K_{1/2}(\lambda) = d\theta_{1/2}(\lambda)/d\lambda$.

From these solutions we obtain the
ground state energy
to first order in $1/L$,
as \cite{ferenc}
\begin{multline}
E_0 = L\epsilon_{\infty} + \epsilon_{d\infty} + \frac{1}{L}\biggl\{
v_{c} \biggl[ \frac{{\Delta 
N_c}^2}{4\xi^2}+\xi^2\Delta_D^2-\frac{1}{12} \biggr]
 \\
+ v_{s} \biggl[ \frac{(\Delta N_{c}-2\Delta N_{s})^2}{4} 
+\frac{1}{2}(\Delta D_{s})^2-\frac{1}{12} \biggr] \biggr\},
\label{eq:Efinal2}
\end{multline}
where $\Delta_D \equiv \Delta D_{c}+\Delta D_{s}$ is the sum of the 
finite-size deviations of $D_c$ and $D_s$ from their
bulk ground state values.
Eq.\ (\ref{eq:Efinal2}) is obtained by an expansion of the ground state 
energy around the thermodynamic limit
$L \rightarrow \infty$ which in our formalism is tantamount to an 
expansion in terms of $(k^\pm \pm k_0)$ and
$(\lambda^\pm \pm \lambda_0)$ around symmetric integration limits
$k_0$ and $\lambda_0$ (cf.\ (\ref{eq:kdef3})).
This is followed by a transformation of variables from $k^\pm$
and $\lambda^\pm$ to
$X_r$ ($r=c,s$ and $X=\Delta N, \Delta D$) evaluated at
$k^\pm = \pm k_0$ and $\lambda^\pm = \pm \lambda_0$, thereby incurring 
as the Jacobian matrix of the transformation a
''dressed charge'' matrix \cite{ferenc} which can be shown to obey the 
same integral equations with the unit matrix as inhomogeneity 
\cite{ferenc}.
In our case, $\xi$ is the function that parameterizes this matrix. 
$v_c$ and $v_s$, finally, are the Fermi velocities in the charge and 
spin sectors.
Note that the leading contribution to the ground--state energy due to 
the magnetic moment
is given by $\epsilon_{d\infty}$.
This contribution has the character of a boundary term \cite{EH91}.
However, the
magnetic moment also affects the parameters in the $1/L$--term, as 
becomes obvious from the decomposition (\ref{eq:rhofinal}) when 
inserted into (\ref{eq:functionsdef}) and (\ref{eq:z0}).

Next, we analyze the finite--size energy to obtain an expression for 
the equilibrium response of the system to an externally applied 
magnetic flux, i.e. the persistent current in the presence of the local 
magnetic moment.
In fact, the persistent current is precisely determined by the 
finite--size contributions proportional to $1/L$ in the energy 
(\ref{eq:Efinal2}), and is obtained by taking the derivative of $E_0$
with respect to the external
Aharonov-Bohm flux. 
Trading the flux $\phi$ for twisted boundary conditions via a gauge 
transformation
leads to an additional shift in the number $D_c$ of electrons 
moved from the left to the
right Fermi points in the charge Fermi sea: $\Delta D_c \rightarrow 
\Delta D_c + \phi$.
Using this replacement in (\ref{eq:Efinal2}), we arrive at the formal 
result for the persistent current:
\begin{equation}
\label{eq:PC}
I(\phi) = - \frac{e\xi^2v_c}{\pi L} 
\bigl[\Delta D_c + \Delta D_s + \phi \bigr].
\end{equation}

Now we are in a position to answer the central question of this 
investigation: How does the presence of the
magnetic moment, which interacts with the electrons in the ring, 
influence the persistent current?
Eq.\ (\ref{eq:PC}) tells us that, to answer this question, we need to 
analyze the effect of the magnetic moment
on the parameters $\Delta D_c$ and $\Delta D_s$.
According to (\ref{eq:functionsdef}) the parts of $\Delta D_c$ and 
$\Delta D_s$ stemming from the magnetic moment
are given by (we ignore bulk terms)
\begin{subequations}
\label{eq:ipara}
\begin{align}
&\Delta D^d_{c}= z^d_{c}(0) +\frac{1}{2} \biggl[ \int_0^{k_0} dk 
+\int_0^{-k_0} d k \biggr] \rho^d_{c}(k)\\
&\Delta D^d_{s}= z^d_{s}(0) + \frac{1}{2} \biggl[ \int_0^{\lambda_0}
d \lambda
+\int_0^{-\lambda_0} d \lambda \biggr] \rho^d_{s}(\lambda),
\end{align}
\end{subequations}
where the density functions $\rho^d_{c}(k)$ and $\rho^d_{s}(\lambda)$ 
are solutions of the integral equations 
with the inhomogeneity $\vec{\rho}^d_0$ (cf.\ (\ref{eq:rho0def})) and 
we have symmetric integration limits $k^\pm = \pm k_0$ and $\lambda^\pm 
= \pm\lambda_0$ in the ground state according to our discussion after 
eq.\ (\ref{eq:Efinal2}).

There is, however, no need to explicitly solve the integral equations 
to obtain further insight into the quantities
$\Delta D^d_c$ and $\Delta D^d_s$.
They follow simply from considering the symmetry of the functions 
involved.
The symmetry properties of all functions derived from the integral 
equations follow from the basic odd symmetry of the BA charge 
rapidities $k$ and the symmetry of the inhomogeneity.
The inhomogeneity $\vec{\rho}_{d0}$ (cf.\ (\ref{eq:rho0def})) is even, 
as are all integral kernels. Further scrutiny 
therefore reveals that the symmetries are such that
$\Delta D^d_c$ and $\Delta D^d_s$ both vanish.
E.g.\ $\vec{\rho}_d$ is an even function in both variables $k$ and 
$\lambda$ and hence, from (\ref{eq:z0}), $z^d_{c}(0) = z^d_{s}(0) = 0$ 
such that, moreover, from (\ref{eq:ipara}), $\Delta D^d_c = 0$ and 
$\Delta D^d_s = 0$.
Hence there is no influence of the magnetic moment
on the persistent current.

We reiterate that our result follows immediately from a 
symmetry analysis of the rapidities and the integral kernels
implied by the BAE (\ref{eq:zdef}). Since these are generic, the result carries over to any model of a quantum impurity coupled to electrons with a dispersion relation that is an even function of momentum, i.e. for example a parabolic (non-relativistic) band, or, if defined on a lattice, a tight-binding band. The reason for this universal behaviour of integrable quantum impurities is that the details of the model do not affect the generic symmetry of the Bethe ansatz, as demonstrated by our analysis above. 
This is also consistent with results obtained for the supersymmetric t-J model, 
where the finite-size ($\propto 1/L$) contribution to the energy due
to twisted boundary conditions was found to be independent of the
presence of an integrable impurity \cite{Bedurftig}.

What is the physics behind this remarkable phenomenon? We propose that 
an answer may be constructed as follows: As is well-known, integrable 
quantum dynamics in one dimension supports only forward scattering 
\cite{IntegrableReview}. It is also known that a forward scattering 
phase shift of a free electron wave function incurred from a local 
static potential has no effect on a persistent current: As was shown by 
Gogolin and Prokof'ev \cite{GP} there is a subtle cancellation (to 
${\cal O}(1/L)$) of contributions to the persistent current from 
phase shifted states, leading to an expression for the current 
in terms of the Fermi level transition amplitude only. Provided that 
the effect of a quantum impurity on the conduction electrons can be 
faithfully encoded by a potential scatterer (much as in Nozi\`ere's local 
Fermi liquid theory of the ordinary Kondo effect \cite{Nozieres}) -- 
{\em and} that this property is not corrupted on a mesoscopic scale -- 
our result would get an elegant and transparent explanation.

From an experimental point of view one may be
concerned that the protection
of the persistent current is not robust against
any deviation from integrability. That is,
any small perturbation would make a difference,
regardless of the perturbation being relevant
or irrelevant in the sense of renormalization group
theory. Thus, to detect a pure protected current
will require ``fine tuning'' of the experimental setup
so as to make sure that the dynamics remains
integrable.

In conclusion, we have demonstrated on quite general grounds that there 
is no influence from a 
quantum impurity on the persistent 
current in a mesoscopic ring when the electron-impurity interaction is 
integrable. We conjecture that this result can be traced back to a 
cancellation of phase shifted contributions to the persistent current, 
in analogy with the simple case of non-interacting electrons in the 
presence of a single forward scattering local potential. To put this 
conjecture on a firm ground, and to extract implications for other 
Aharonov-Bohm (or Aharonov-Casher \cite{EJS}) geometries, is an 
interesting and challenging problem.

{\bf Acknowledgments} \ We thank I. Affleck, N. Andrei, and A. 
Zvyagin for discussions that prompted us to undertake this 
investigation. We are also grateful to S. Eggert and C. Stafford for 
valuable discussions. H.J. acknowledges support from the Swedish 
Research Council.


\end{document}